# Assessing Foundational Medical 'Segment Anything' (Med-SAM1, Med-SAM2) Deep Learning Models for Left Atrial Segmentation in 3D LGE MRI

Mehri Mehrnia
Department of Biomedical Engineering
Northwestern University, Chicago, USA
mehri.mehrnia@northwestern.edu

Mohamed Elbayumi
Department of Biomedical Engineering
Northwestern University, Chicago, USA
mohamedelbayumi2027@u.northwestern.edu

Mohammed S. M. Elbaz
Feinberg School of Medicine
Northwestern University, Chicago, USA
mohammed.elbaz@northwestern.edu

## ABSTRACT

Atrial fibrillation (AF), the most common cardiac arrhythmia, is associated with heart failure and stroke. Accurate segmentation of the left atrium (LA) in 3D late gadolinium-enhanced (LGE) MRI is helpful for evaluating AF, as fibrotic remodeling in the LA myocardium contributes to arrhythmia and serves as a key determinant of therapeutic strategies. However, manual LA segmentation is labor-intensive and challenging. Recent foundational deep learning models, such as the Segment Anything Model (SAM), pre-trained on diverse datasets, have demonstrated promise in generic segmentation tasks. MedSAM, a fine-tuned version of SAM for medical applications, enables efficient, zero-shot segmentation without domain-specific training. Despite the potential of MedSAM model, it has not yet been evaluated for the complex task of LA segmentation in 3D LGE-MRI. This study aims to (1) evaluate the performance of MedSAM in automating LA segmentation, (2) compare the performance of the MedSAM2 model, which uses a single prompt with automated tracking, with the MedSAM1 model, which requires separate prompt for each slice, and (3) analyze the performance of MedSAM1 in terms of Dice score(i.e., segmentation accuracy) by varying the size and location of the box prompt.

## Introduction

Atrial fibrillation (AF) is the most common cardiac arrhythmia, affecting nearly 50 million people worldwide and contributing to severe comorbidities such as heart failure, stroke, and increased mortality rates (1). The progression of AF is associated with fibrosis in the left atrium (LA)—the upper left heart chamber—which disrupts normal cardiac rhythm and exacerbates AF symptoms. Recent advancements in 3D late gadolinium-enhanced magnetic resonance imaging (3D LGE-MRI) have enabled noninvasive assessments of LA fibrosis and scar tissue, providing detailed insights into native fibrosis and post-ablation scarring, which is crucial for effective treatment planning (2, 3). Manual LA segmentation is often required as a prerequisite for analysis or quantification. However, 3D LGE MRI is a high resolution scan that results in 30-50 slices per scan making it tedious and time-consuming task and inter/intra-observer segmentation variability can hinder reproducibility that is critical for clinical evaluation (4). Anatomical complexity with variability in geometry and size of LA structures, such as the pulmonary veins and left atrial appendage, further complicates segmentation across different patients (5, 6).

Deep learning techniques have significantly advanced medical image segmentation, offering the potential to automate traditionally laborious tasks and enabling large-scale data analysis (4). These models have demonstrated the ability to learn complex image features, accurately identifying specific anatomical structures and pathological regions. However, the success of these deep learning models often hinges on the availability of large and diverse domain-specific datasets for training, which poses a significant challenge. The diversity in imaging protocols, scanner types, and patient populations makes it challenging for such domain-specific-trained models to generalize effectively across different conditions(7, 8). Recently, the Segment-Anything Model (SAM) was introduced as a foundational segmentation model, trained on a massive dataset of one billion images (9). SAM stands out for its ability to generalize across different domain vision tasks employing user-defined prompts such as points or bounding boxes. Fine-tuning SAM for medical imaging, MedSAM, has shown improved performance over specialist domain-specific-trained convolutional neural networks (CNNs), with ability to adapt to medical image data from different modalities without the need for re-training (10–12). However, these generic foundational MedSAM models have not yet been tested for the challenging LA segmentation from 3D LGE-MRI, where its generalization capabilities and prompt-based input could facilitate LA segmentation. This study seeks to assess MedSAM's potential for LA segmentation by evaluating two approaches: MedSAM1, a 2D model that requires box prompt per each slice, and MedSAM2, a more recent 3D model that uses a single prompt with automated tracking across slices thus requiring one prompt per scan. By evaluating the efficiency and accuracy of these models, this study explores whether foundational segmentation models can address the longstanding challenge of LA segmentation in LGE-MRI, thereby reducing manual effort towards enhancing both consistency and efficiency in MRI evaluation of AF patients. Therefore, the aims of this study were to 1) assesses and compare MedSAM1 and MEDSAM2 performance for prompt-based LA segmentation from 3D LGE MRI of AF patients. 2) Conduct a sensitivity analysis on the impact of box prompt size and location on MedSAM1's segmentation performance.

## Materials and Methods

### Ethics Statement

The data were sourced from the publicly available CARMA Utah database(13).

### Study Population

The cohort consisted of 44 pre-ablation AF patients (mean age 67 ± 9 years, 34% female) obtained from the publicly available CARMA 3D LGE-MRI database(13). This dataset includes a total of 1227 2D slices with manual segmentations, provided in the public dataset, of both the (1) endocardial surface of the LA, including the pulmonary vein antrum regions, and (2) the LA wall, excluding the pulmonary vein antrum, mitral valve, and left atrial appendage.

### MRI Protocol

The 3D LA LGE MRI scans were acquired on either a 1.5 Tesla Avanto or a 3.0 Tesla Verio clinical scanner (Siemens Medical Solutions, Erlangen, Germany). The scans were performed 15 minutes after contrast agent injection (0.1 mmol/kg, Multihance [Bracco Diagnostic Inc., Princeton, NJ]) using a 3D inversion recovery, respiration-navigated, ECG-gated, gradient echo pulse sequence. The typical acquisition parameters were: free-breathing with navigator gating, a transverse imaging volume with two different sizes: $576 \times 576 \times 44$ voxels and $640 \times 640 \times 44$ voxels but with the same voxel size of 1.25 * 1.25 * 2.5 mm3 (reconstructed to $0.625 \times 0.625 \times 2.5$ mm3), repetition time (TR)/echo time (TE) of 5.4/2.3 ms, flip angle of 20 degrees, and inversion time (TI) of 270-310 ms. GRAPPA with R = 2 and 46 reference lines was used. ECG gating was employed to acquire phase encoding views during the diastolic phase of the LA cardiac cycle, with the time interval between the R-peak of the ECG and the start of data acquisition defined by cine imaging of the LA. The TI value was determined using a scout scan (15). LA was covered by an average of 30 slices per 3D LGE MRI scan.

### Med-SAM LA Segmentation pipeline

We evaluated the performance of the foundational MedSAM1 and MedSAM2 models for semi-automated,

prompt-based LA segmentation, using the analysis pipeline illustrated in Fig. 1. For both MedSAM1 and MedSAM2, no preprocessing was applied to the intensity or dimension of 3D LGE-MRIs.

*Med-SAM1 segmentation pipeline*

MedSAM1 operates as a 2D segmentation model, requiring a user-defined box prompt for each individual LA slice, Fig. 1a (11). The steps in the segmentation pipeline are: 1) Input: The 3D LGE-MRI dataset is loaded into the system. 2) Box Prompt Definition: For each slice, a rectangular box prompt is drawn to encompass the LA and pulmonary veins (PVs). The box dimensions were derived from ground truth segmentations and expanded by 10% to simulate a natural human prompt approach. 3) Saving Prompt Coordinates: The coordinates of each box prompt are saved for subsequent analysis, including evaluation of sensitivity to changes in prompt size. 4) Segmentation: MedSAM1 segments each 2D slice based on the provided box prompts, and the segmented slices are saved individually. 5) Volume Reconstruction: The segmented 2D slices are compiled to reconstruct the complete 3D LA volume for further analysis. Segmentation was performed using the code in (14).

*Med-SAM2 segmentation pipeline*

MedSAM2 operates as a 3D segmentation model using a single freehand (scribble) prompt on one slice followed by automated tracking to segment the subsequent slices, Fig. 1b(15). The analysis pipeline includes the following steps: 1) Slice Selection and Preparation: For each patient, the slices containing the LA (including the PVs) were selected from the original set of 44 slices. These selected slices were compiled into a video format (AVI clip), with each frame representing a slice, using MATLAB to ensure compatibility with the MedSAM2 platform(16).2) Scribble Prompt: The user applies a freehand scribble on the initial LA slice as displayed in the video. 3) Automated Tracking: MedSAM2 segments the scribbled slice and automatically tracks and segments the LA in the subsequent slices. 4) Segmentation Mask Generation: Segmentation results are outputted as a video file and as individual 2D segmentation masks for each slice. Slices that could not be segmented by MedSAM2 are saved as blank images. 5) 3D Volume Reconstruction: The 2D segmentation masks are aggregated to reconstruct the complete 3D LA volume, using the original spatial resolution and orientation metadata from the MRI data to ensure accurate 3D reconstruction.

*Evaluation metric*

To assess segmentation accuracy, we calculated both 2D and 3D Dice scores(17), as well as 3D Hausdorff Distance (HD)(18) and Average Symmetric Surface Distance (ASSD)(19) for each patient. Both HD and ASSD are reported in millimeters (mm). These metrics provided a detailed evaluation of the model's performance in accurately delineating anatomical regions across both individual image slices and the full reconstructed volume. The 2D Dice score was calculated for each slice, with the ground truth defined as the region of interest encompassing both the LA wall and the endocardium (blood pool). The 3D Dice score was computed based on the entire LA volume, reconstructed by stacking the 2D slices. We reported 2D Dice scores for a total of 1,227 slices from 44 patients, along with volume-based 3D Dice scores for all 44 AF patients. The processing time, including prompt creation and segmentation, for each model was recorded to assess efficiency. To ensure a fair and direct comparison between the two models, we accounted for the missing slices in MedSAM2 by filtering and including only the corresponding slices from MedSAM1, based on the available slices in MedSAM2. Additionally, the left atrial (LA) volume was calculated using the voxel size of the MRI scans and compared to the ground truth to assess the accuracy of segmentation in cubic millimeters ($mm^3$) for both models.

*Sensitivity Analysis*

We conducted a sensitivity analysis to assess the effect of two key parameters on segmentation performance in the MedSAM1 model: box prompt size and box prompt location. Our goal was to evaluate how variations in these parameters influence segmentation accuracy of the tested foundational models. We performed sensitivity analysis to the two key sources for variability as follows:

1) *Sensitivity to Box Prompt Size.* To assess the effect of box prompt size on segmentation accuracy, we incrementally increased the box size by 10% at each of four steps, resulting in 10%, 20%, 30%, and 40% increases relative to the original size. For each box prompt size, the left atrium (LA) was segmented using the

MedSAM1 model, and segmentation quality was evaluated using the Dice score, comparing predicted segmentations to the ground truth. Additionally, agreement and variability were analyzed using ICC and CoV between the initial Dice score and those with enlarged box prompts.

*2) Sensitivity to Box Prompt Location.* To simulate potential observer variability in defining the box prompt, we introduced random shifts to the box location. The centroid of the box was randomly moved within 10% of the original prompt size in both the x and y directions, shifting the box independently per slice. Segmentation accuracy was evaluated using the Dice score, comparing the results to the ground truth.

## *Statistical Analysis*

Continuous variables were presented as mean ± standard deviation (SD) or median [interquartile range, IQR] with IQR presented as [25th, 75th percentile], depending on the data distribution as determined by the Shapiro-Wilk test. For sensitivity analyses, a paired Wilcoxon signed-rank test was applied to assess the statistical significance of changes in Dice scores due to variations in box size and location. The intraclass correlation coefficient (ICC) and coefficient of variation (CoV) and Bland Altman analysis were employed to assess the agreement between left atrial (LA) volume estimates from MedSAM1 and MedSAM2 and the ground truth. Additionally, the correlation between the LA volume obtained from MedSAM1/MedSAM2 and the ground truth was evaluated using the Spearman correlation coefficient ($\rho$).

## **Results**

## *Med-SAM 1 Segmentation performance*

MedSAM1 achieved a 2D Dice score of 0.84 [0.73, 0.89], based on comparisons between the predicted segmentation and the ground truth for 1,227 slices, with an average of 30 slices per patient (Fig. 3.a). For 3D segmentation, MedSAM1 attained a Dice score of 0.81 ± 0.05 when comparing the entire left atrium (LA) volume with the ground truth (N=44), Fig. 3.b. The Bland-Altman plot in Fig. 4a represents the sensitivity analysis of LA volume estimated by MedSAM1 compared to the ground truth, showing good agreement (ICC = 0.77, CoV = 7.87%).

The volume-based segmentation accuracy, assessed using HD and ASSD metrics, showed values of 19.96 [18.12, 22.82] mm for HD and 2.73 ± 0.7 mm for ASSD in MedSAM1, Tabel 1.

Figure 5 presents three examples of MedSAM1's segmentation performance, shown through 2D Dice scores, ranging from high accuracy (0.87) to average (0.79) to lower accuracy (0.70). In the example with high dice score(Fig 5.a), the left atrial (LA) wall and pulmonary veins (PVs) are fully and accurately segmented. However, as performance declines, portions of the LA wall are mis-segmented, and the aorta is incorrectly included as part of the target segmentation. In MedSAM1, creating each box prompt takes approximately 10 seconds, resulting in a total processing time of about 5 minutes per scan (average 30 slices per scan).

*Med-SAM 2 Segmentation performance*

MedSAM2 achieved a 2D Dice score of 0.84 [0.64, 0.91] (Fig. 3a). Segmentation of the full 3D LA volume using MedSAM2 resulted in a 3D Dice score of 0.81 ± 0.05 (N=44) (Fig. 3b). After the scribble prompt is applied over a slice with at least partial LA visibility, MedSAM2 typically performs forward auto-tracking. On average, MedSAM2 missed approximately 2 slices per scan during segmentation across the cohort of 44 patients. However, if the scribble is applied at the very beginning of the LA, it can cause significant segmentation errors across the entire LA segmentation.

The Bland Altman plot in Fig. 4. b represents an agreement analysis of LA volume estimated by medSAM2 compared to the ground truth. The agreement between these volumes was good (ICC = 0.70, CoV = 11.50%). Volume-based segmentation accuracy, assessed using HD and ASSD metrics, showed values of 29.89 ± 8.78 mm for HD and 3.86 ± 1.87 mm for ASSD in MedSAM2, Tabel 1.

The LA volumes (mm³) estimated by MedSAM1 and MedSAM2 show a significant correlation (Spearman's rho = 0.72, p < 0.001), with good agreement (ICC = 0.74) and relatively low variability (CoV = 7.79), as shown in Fig. 4c. The paired comparison further demonstrates no significant difference between the volumes estimated by MedSAM1 and MedSAM2 (p = 0.08).

Figure 6 presents three examples of MedSAM2's segmentation performance, depicted through 2D Dice scores ranging from high accuracy (0.85) to average (0.74) to low accuracy (0.58). In the example with high score (Fig 6.a), the LA wall and pulmonary veins (PVs) are fully identified. However, as the Dice score declines, portions of the LA are mis-segmented, and the aorta is incorrectly identified as the LA.

MedSAM2 requires only a single scribble prompt to segment the entire 3D LGE-MRI volume in approximately 20 seconds.

Fig.7-8 provides examples where MedSAM1 outperformed MedSAM 2 and vice versa as compared to the ground truth. Fig. 7 shows an example case where MedSAM1(dice=0.77) has higher dice score than MedSAM2 (Dice=0.55), which struggled with misidentifying the aorta as part of the target while attempting to auto-track the LA in slices. Fig. 8 presents another example where MedSAM2 (dice=0.84) achieved higher dice score than MedSAM1(dice=0.77). In Fig. 8 MedSAM1 failed to accurately segment the LA wall, partially included the aorta, and showed difficulty in correctly segmenting the PVs.

### *Sensitivity Analysis of Box Prompt Size and location in MedSAM1*

### *Sensitivity to Box Prompt Size*

Sensitivity analysis demonstrated a decrease in 2D Dice scores with increasing box prompt sizes, where 10%, 20%, 30%, and 40% increases in prompt size resulted in Dice scores of 0.78 [0.69, 0.84], 0.69 [0.59, 0.75], 0.59 [0.50, 0.66], and 0.50 [0.42, 0.58], respectively, over N=1227 slices. Each 10% increase in box prompt size corresponded to an approximate 10% reduction in segmentation accuracy.

A comparison between the Dice scores of segmentation results for each enlarged box prompts and the original box prompt demonstrated that box prompt size has a significant impact on the accuracy of MedSAM1 (p < 0.001).

A comparative analysis of increasing the prompt size by 10%, 20%, 30%, and 40% revealed a decline in ICC from 0.89 to 0.62, 0.36, and 0.21, respectively, and an increase in CoV from 3.75 to 8.07, 13.04, and 18.22

(Table 2).

*Sensitivity to Box Prompt location*

Variability in the box prompt location resulted in a 2D Dice score of 0.82 [0.71, 0.88], indicating a negligible impact on segmentation performance compared to the original dice of 0.84 [0.73, 0.89], Fig.10. The results of the agreement and variability analysis, with an ICC of 0.96 and CoV of 3.20%, further corroborate the negligible impact of box location variability on segmentation accuracy, Table 2.

**Discussion**

Our study evaluated the performance of the MedSAM1 and MedSAM2 models for segmenting the left atrium (LA) in 3D LGE-MRI and conducted a sensitivity analysis of MedSAM1 to assess the impact of prompt size and location on segmentation accuracy. Both models demonstrated promising results, with comparable performance in both 2D and 3D volumetric segmentation. MedSAM2's capability to segment the entire LA volume with a single prompt significantly enhanced its processing time efficiency, while MedSAM1, though with comparable Dice score, required significantly more manual effort and was more time-consuming due to the need for individual prompts for each 2D slice. In 3D evaluation of segmentation, MedSAM1 negligibly MedSAM2 in both HD: 19.96 mm vs. 29.87 mm and ASSD: 2.73 mm vs. 3.86 mm. The performance of MedSAM2 was significantly affected by unsuccessful auto-tracking, leading to the incorrect segmentation of the aorta (AO) instead of the left atrium (LA), as observed in the outliers of the box plot representation.

Segmentation volume (mm³) comparisons between MedSAM1 and MedSAM2, when evaluated against the ground truth, demonstrated similar performance. MedSAM1 achieved an ICC of 0.77 and a CoV of 7.87%, while MedSAM2 showed an ICC of 0.70 and a CoV of 11.50%, indicating comparable accuracy. This similarity was further supported by a direct comparison of the segmentation volumes between MedSAM1 and MedSAM2, with a Spearman's rho of 0.72, ICC of 0.74, and CoV of 7.79%.

Compared to other methods in the literature, such as U-Net-based CNNs specifically trained on 3D LGE MRI

data, which achieve Dice scores as high as 0.93 and an ASSD of 0.7 mm (14–16), MedSAM models exhibit slightly lower segmentation performance. However, these task-specific models require extensive dataset-specific training and often rely on manual corrections, limiting their generalizability across different datasets and imaging modalities.

In contrast, the MedSAM models are generic foundational models that do not require retraining specifically on 3D LGE MRI data, making them more versatile and adaptable for various segmentation tasks.

The sensitivity analysis provides valuable insights into the potential and limitations of foundational models for automated segmentation of complex anatomical structures. It revealed that prompt size significantly impacts MedSAM1's segmentation accuracy, with a roughly 10% decrease in Dice score as larger prompts capture unnecessary surrounding structures, reducing precision. This highlights the importance of selecting an optimal prompt size for effective segmentation, especially in anatomically complex regions. In contrast, prompt location had a minimal effect on segmentation accuracy.

However, a notable challenge remains in accurately segmenting the pulmonary vein (PV) antrum in left atrium (LA) segmentation, a crucial factor for atrial fibrillation (AF) therapy and ablation planning. The PV antrum contains electrically active tissue that plays a key role in understanding AF and predicting radiofrequency (RF) ablation outcomes. In manual segmentations, the extent of PV inclusion is often subjectively determined by experts based on various anatomical cues, such as the beginning of PV branching, where the PV stops narrowing, or a set distance (e.g., 10 mm or approximately three times the LA wall thickness) (13). The inability of MedSAM models to inherently capture these nuanced anatomical considerations often results in over- or under-inclusion of PVs, depending on how the prompts are defined. This limitation is particularly evident in MedSAM1, which frequently struggled around the PV antrum due to the use of per-slice box prompts that encompassed multiple structures, leading to errors in segmentation. MedSAM1 is designed for 2D data segmentation, which makes it less suitable for 3D LGE-MRI segmentation tasks. In the pulmonary vein (PV) antrum areas, LA can be presented as multiple isolated structures (depending on the slice orientation), yet we consistently used a single box prompt with a single output, which might have contributed to inaccurate segmentations and future studies should assess whether returning multiple structures could resolve this issue. MedSAM2, while more efficient in handling large-scale segmentations, also faced difficulties in low-contrast areas and often over-segmented nearby structures like the aorta, further complicating PV antrum delineation. These findings align with previous studies showing that SAM models perform well on tasks with clear boundaries but struggle with targets that have weak boundaries or low contrast (20, 21).

Despite the slightly lower performance, the MedSAM models show promise due to their ability to achieve reasonable segmentation accuracy without the need for problem-specific training. This flexibility, combined with the efficiency of MedSAM2, positions these models as valuable tools in clinical settings where generalizability and speed are prioritized. However, the challenge of accurately capturing complex anatomical regions like the PV antrum underscores the need for continued refinement, particularly in addressing the limitations of automatic segmentation in medically significant structures.

*Limitations*

The public dataset used in this study is limited, comprising data from a single vendor and one center. Additionally, MedSAM1 requires a box prompt for each slice, which can be a significant burden when dealing with large datasets. MedSAM2 used herein employs only forward auto-tracking of the LA across subsequent slices after applying a scribble prompt. However, this often leads to missing segmentation on the initial slices where LA is small, or contrast is suboptimal. If the scribble is placed at the very start of the LA to address this issue, the entire LA segmentation may be compromised, as the model struggles to recognize the LA from a minimal scribble at the start. Nevertheless, since MedSAM2 requires only a single scribble for the entire 3D LGE-MRI, it may offer a more efficient approach for large datasets. Sensitivity analysis was only performed for MedSAM1, while it was not conducted for MedSAM2 due to limitations in the off-shelf online system used which did not allow a systematic approach to test such impact thus there remains a gap in the understanding of prompt variability effects on MedSAM2. The box prompt size was increased up to 40% because, in the absence of negative prompts, expanding the box prompt to include surrounding structures, such as the aorta, could result in significant segmentation errors. It is also important to recognize that segmentation performance can vary depending on differences in imaging modalities, patient populations, and acquisition methods, even for the same segmentation task. However, these factors were not assessed herein due to limitations of retrospective analysis of the available public datasets and future studies are warranted. Future work should assess whether employing multiple box prompts could help prevent missegmentation in the pulmonary vein (PV) antrum areas for MedSAM1. For MedSAM2, automating the scribble process could decrease segmentation time. Additionally, incorporating point prompts and using negative prompts for the aorta could reduce frequent missegmentation of the aorta as part of the target region.

**Conclusions**

This study demonstrates the potential of foundational models, MedSAM1 and MedSAM2, for automating the challenging task of LA segmentation in 3D LGE-MRI scans of atrial fibrillation (AF) patients using simple user prompts. Both models exhibited promising performance, with MedSAM2 achieving comparable accuracy to MedSAM1 while being more time and labor-efficient due to the need for only a single prompt per scan. This efficiency may make MedSAM2 potentially helpful for large-scale or time-sensitive clinical workflows. However, both models faced challenges in accurately segmenting LA in the presence of other structures such as the pulmonary veins within the prompt or in the presence of low-contrast regions, highlighting the need for further future work.

**Data Availability**

Data is publicly available.

# Tables

Table 1: Result of volume (3D) analysis of MedSAM1 and MedSAM2

|  | Dice↑ | Hausdorff Distance (mm)↓ | Average Symmetric Surface Distance(mm)↓ |
|---|---|---|---|
| MedSAM1 | 0.81 ± 0.05 | 19.96 [18.12,22.82] | 2.73 ± 0.77 |
| MedSAM2 | 0.80 [0.75,0.85] | 29.89 ± 8.78 | 3.86 ± 1.87 |

Legend: The table presents the 3D segmentation performance of MedSAM1 and MedSAM2, comparing metrics such as Dice score (higher is better), Hausdorff Distance (HD, in millimeters, lower is better), and Average Symmetric Surface Distance (ASSD, in millimeters, lower is better).

Table 2: Sensitivity Analysis Results of Box Prompt Size and Location on MedSAM1 Segmentation Accuracy (dice score), N=1227

| MedSAM1 | box prompt size increased by 10% | box prompt size increased by 20% | box prompt size increased by 30% | box prompt size increased by 40% | box prompt with location variability |
|---|---|---|---|---|---|
| IntraClassCorrelation (ICC)↑ | 0.89 | 0.62 | 0.36 | 0.21 | 0.96 |
| Coefficient of Variation (CoV)%↓ | 3.75 | 8.07 | 13.04 | 18.22 | 3.20 |

Legend: This table presents the results of the sensitivity analysis for MedSAM1 segmentation accuracy with varying box prompt sizes and location variability. The analysis includes metrics such as IntraClassCorrelation (ICC, higher is better for consistency) and Coefficient of Variation (CoV, lower is better for variability). As the box prompt size increases from 10% to 40%, the ICC declines, indicating reduced agreement with the ground truth, while CoV increases, indicating greater variability. In contrast, prompt location variability had minimal impact, with a high ICC of 0.96 and low CoV of 3.20%.

# Figures

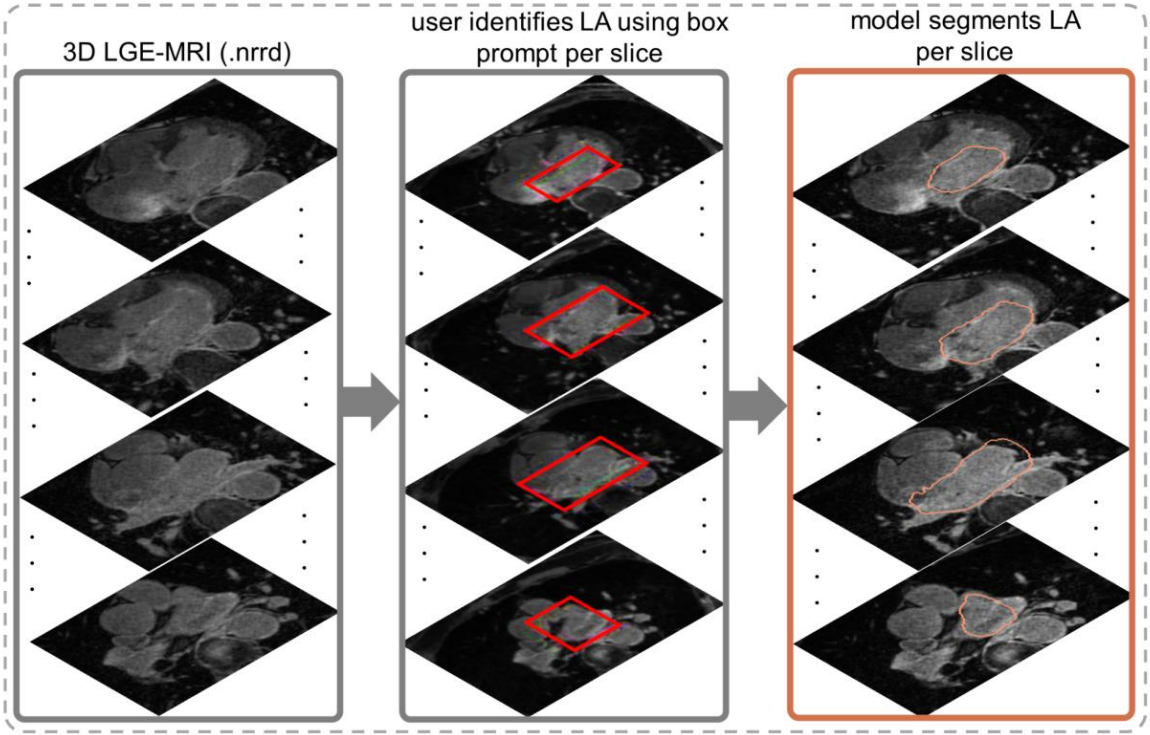

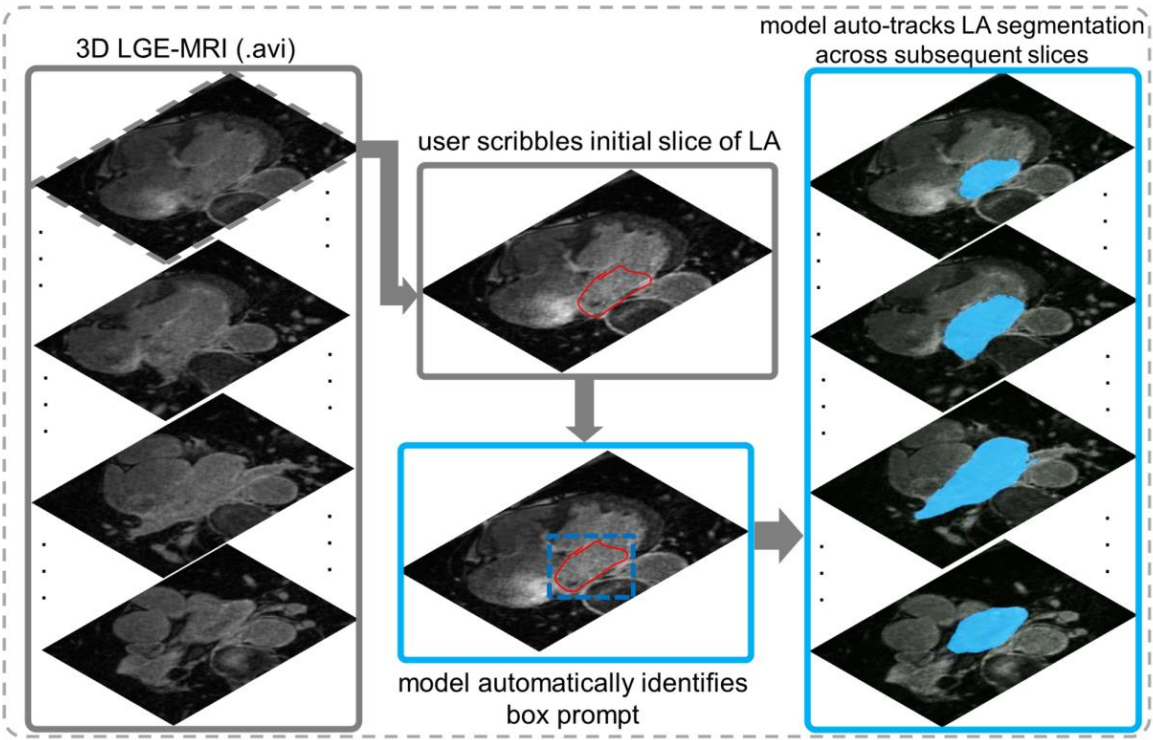

Figure 1: LA segmentation pipeline for MedSAM1 and MedSAM2 models. (a) MedSAM1 pipeline: This process involves applying a user-defined box prompt to each 2D slice of the left atrium (LA). Each slice requires a separate prompt, and segmentation is carried out on a slice-by-slice basis. (b) MedSAM2 pipeline: MedSAM2 utilizes a single freehand (scribble) prompt placed on one slice at the beginning of the LA. The model then automatically tracks the LA across subsequent slices to generate a full 3D segmentation of the LA volume. Note: The key difference between the two pipelines lies in the number of prompts required: MedSAM1 requires a prompt for each individual slice, whereas MedSAM2 needs only a single prompt to generate the entire volume.

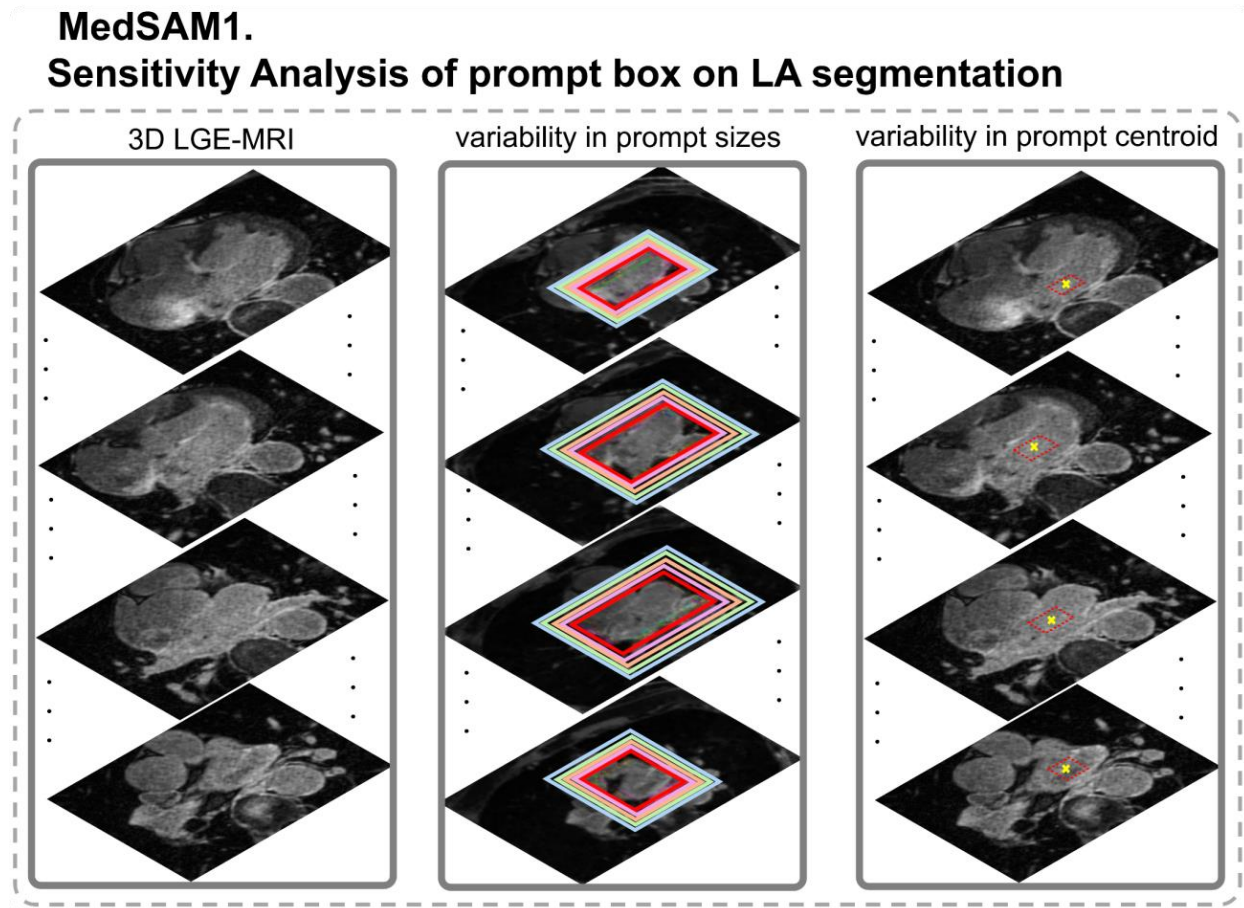

Figure 2. Illustration of the experimental setup for sensitivity analysis of MedSAM1, demonstrating variations in box prompt sizes and locations for left atrial segmentation. In the second column, each box, shown in a different color, was increased by 10%, 20%, 30%, and 40% relative to the initial prompt size. In the third column, location variability is visualized by dashed areas, representing a 10% shift margin around the center of the original prompt box. The centroid of the box prompt can move within this defined dashed area.

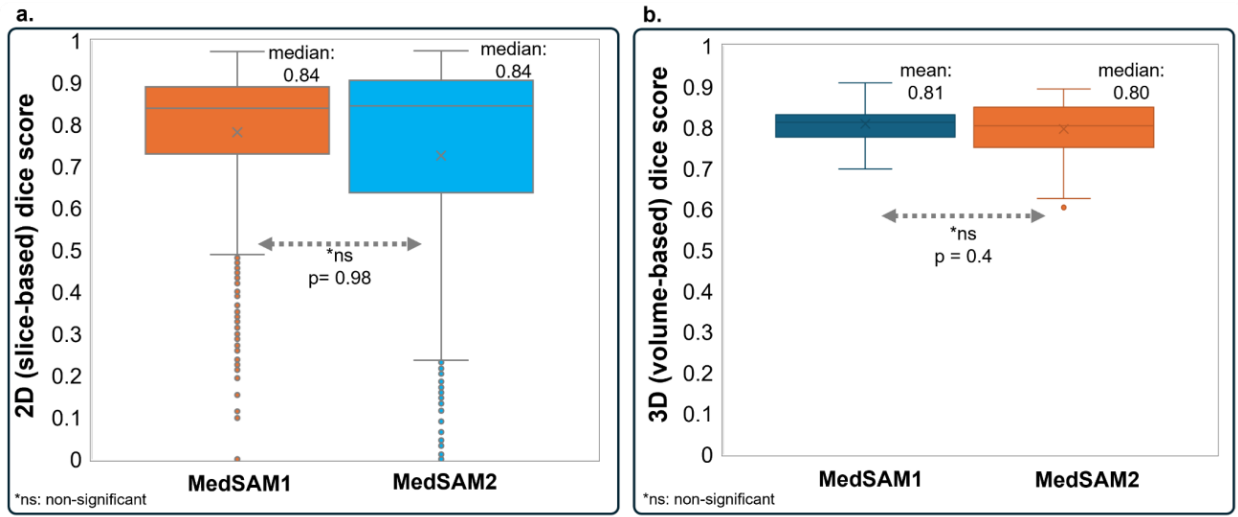

Figure 3.a. box plot showing the segmentation performance of MedSAM1 and MedSAM2, based on N= 1,227 2D (slice-based) Dice scores from 44 patients. Circles represent outlier Dice scores, and the large 'X' marks the mean Dice score for each model. The results indicate no significant difference between the two models, b. 3D Dice score comparison between MedSAM1 and MedSAM2, N= 44 AF patients.

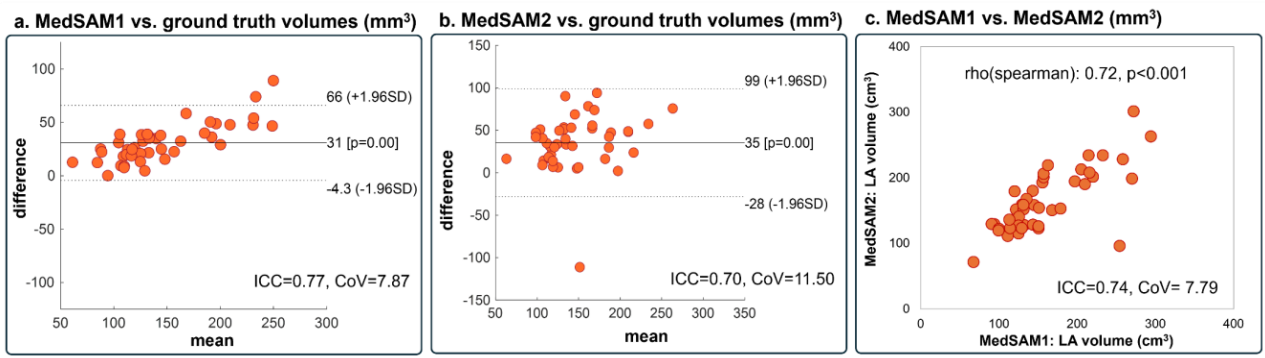

Figure 4. a, b. Bland-Altman plots comparing LA volume estimation from MedSAM1 and MedSAM2 with ground truth, respectively. The x-axis represents the average LA volume of MedSAM1/MedSAM2 and ground truth, while the y-axis represents the difference between the estimated LA volume from MedSAM1/MedSAM2 and the ground truth. c. Scatter Plot of LA Volume (cm³) from MedSAM1 and MedSAM2. Each point represents a single patient, N=44.

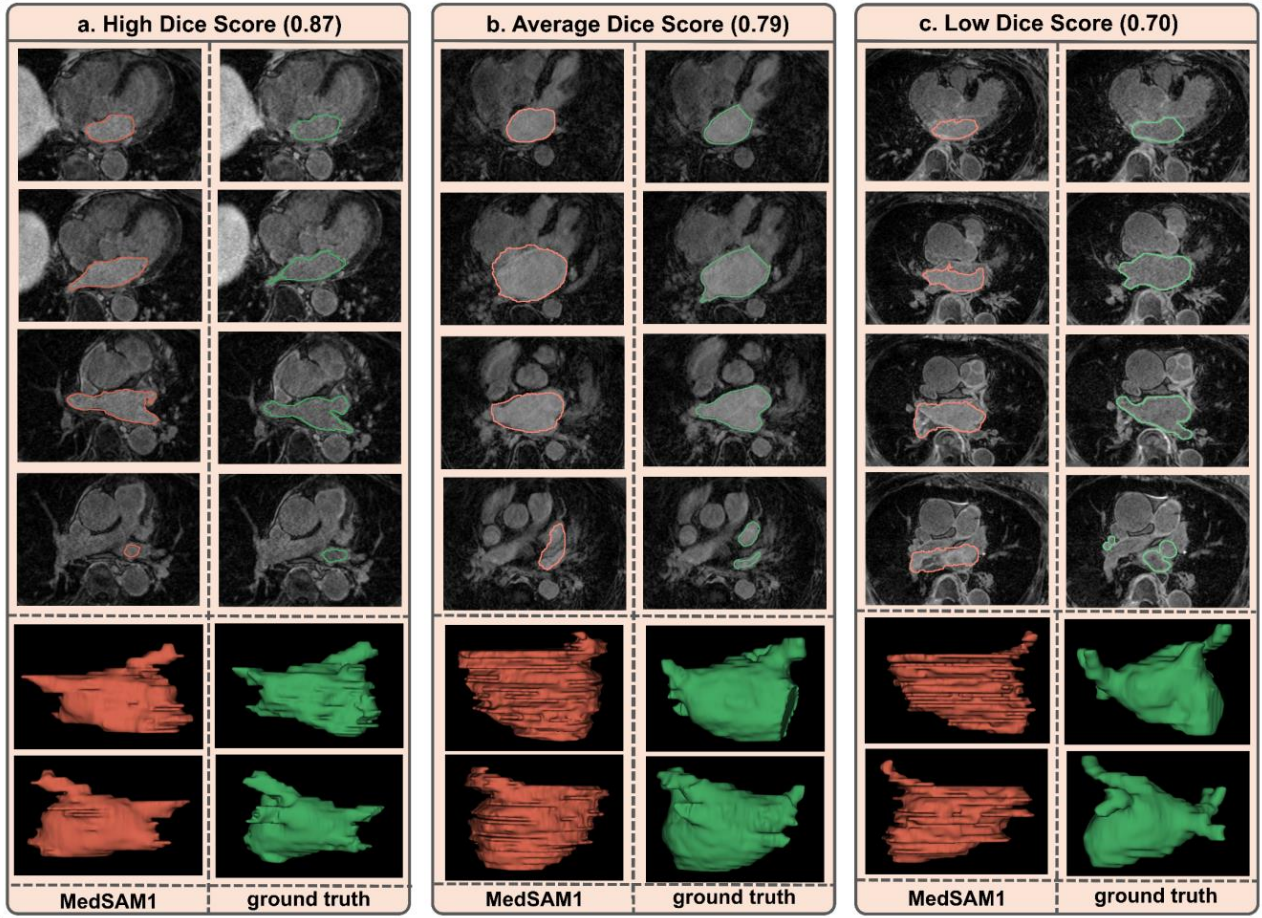

Figure 5. MedSAM1 segmentation examples with varying performance: (a) High Dice score, (b) Average Dice score, and (c) Low Dice score. For each example, the 2D and 3D segmentation results (first column) are shown alongside the ground truth (second column). (a) The model segments the entire slice with high quality, accurately identifying the pulmonary veins. (b) In areas with blurry borders, the segmentation exceeds the LA wall. (c) The LA wall is under-segmented, and the aorta is incorrectly segmented along with the pulmonary veins. When multiple pulmonary vein antrums are present, the use of a single box misleads the model, resulting in incorrect segmentation of surrounding areas. The Dice score represents the average Dice value across all slices of the patient.

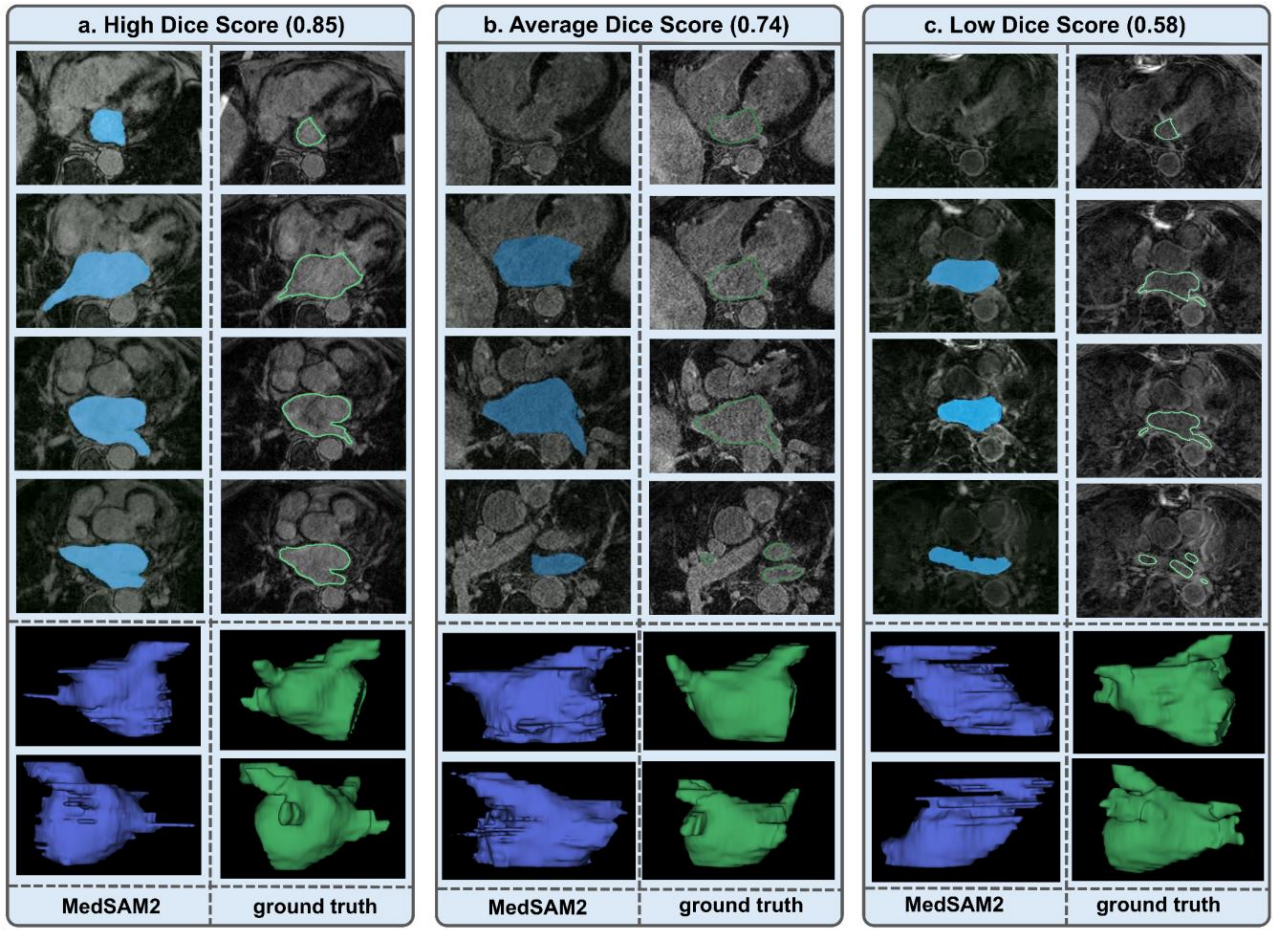

Figure 6. MedSAM2 segmentation examples with varying performance: (a) High Dice score, (b) Average Dice score, and (c) Low Dice score. For each example, the 2D and 3D segmentation results (first column) are shown alongside the ground truth (second column). (a) The model segments the entire slice with high quality, accurately identifying the LA wall and pulmonary veins. (b) The initial slice is missing, and in areas with blurry borders, the segmentation exceeds the LA wall. (c) The initial slice is missing, the LA wall is misrepresented, and the aorta is incorrectly tracked instead of the LA. The Dice score represents the average Dice value across all slices of the patient.

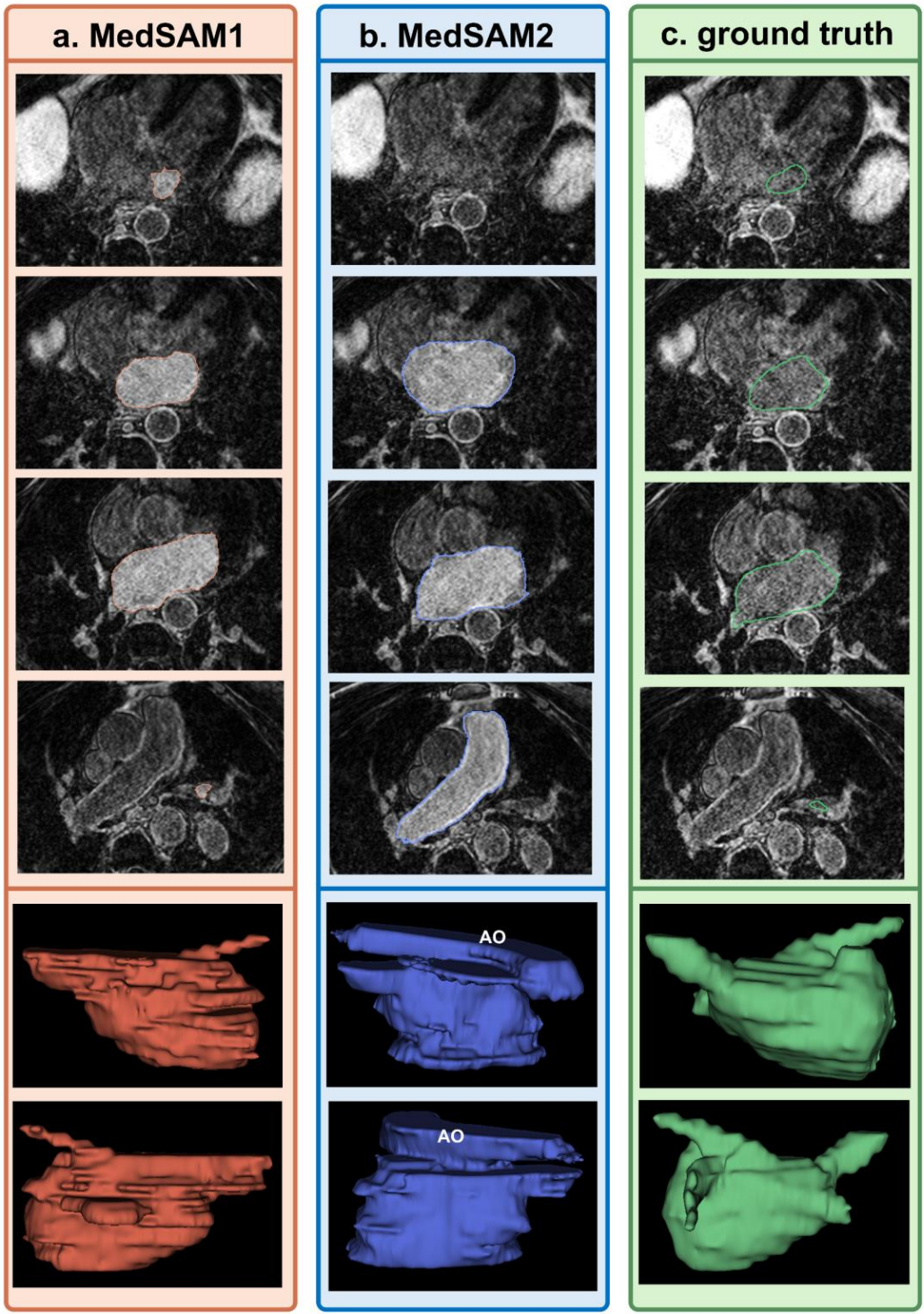

Figure 7. An example of segmentation results with both slice- and volume-based representations from MedSAM models alongside the ground truth. (a) MedSAM1 (Dice = 0.77) outperforms MedSAM2 (Dice = 0.55) in segmentation accuracy.

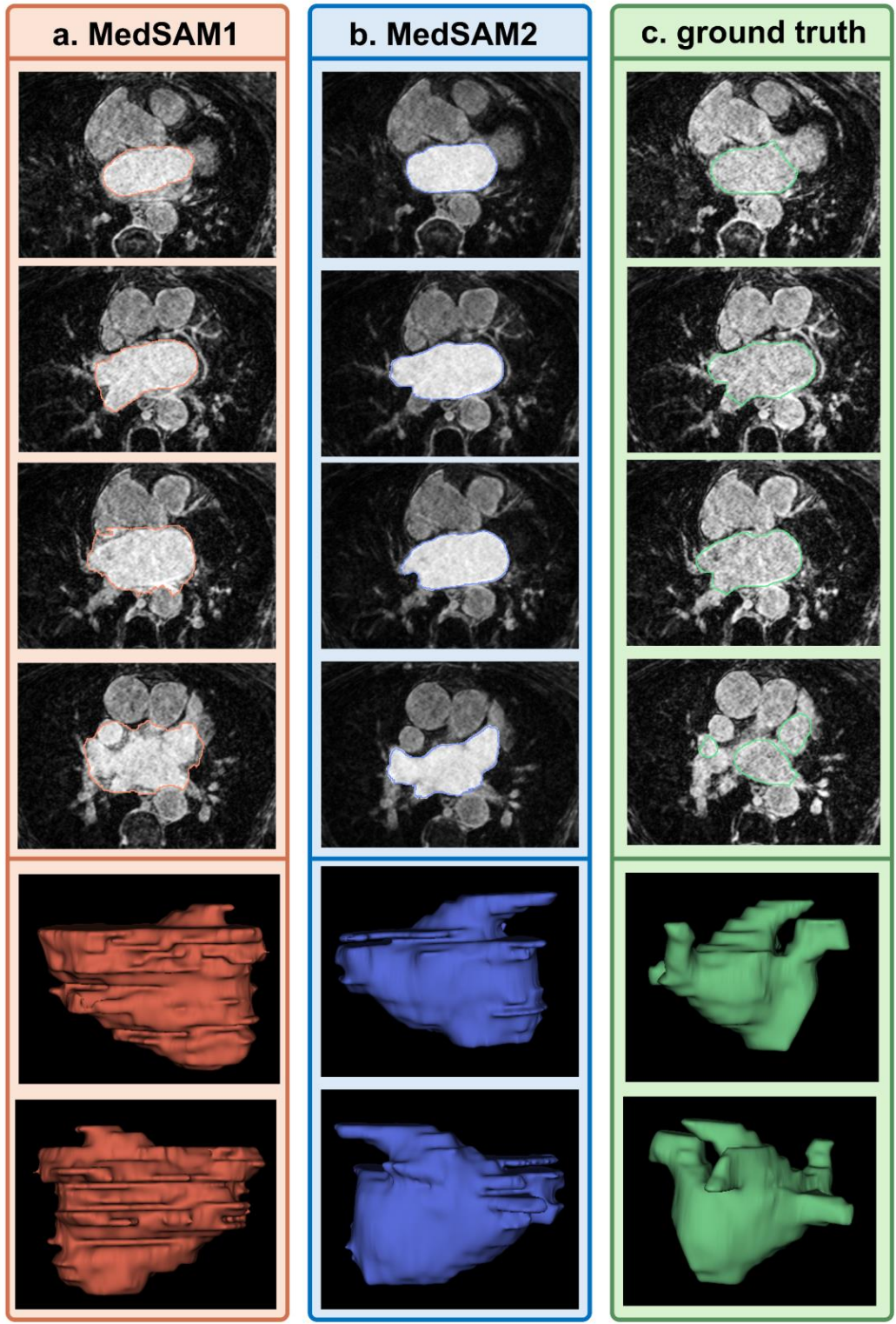

Figure 8. An example of segmentation results with both slice- and volume-based representations from MedSAM models along with ground truth. MedSAM2(dice = 0.84) outperforms MedSAM1(0.77).

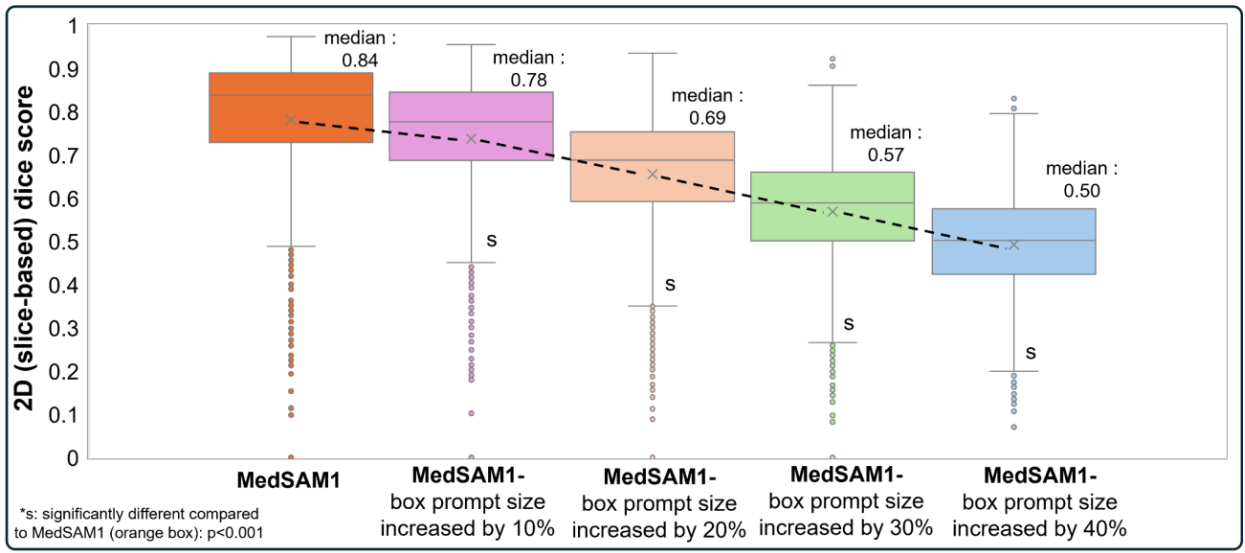

Figure 9. MedSAM1 - Box Plot of Dice Scores for Varying Box Prompt Sizes. From left to right, the first box represents the 2D Dice score with the original box prompt size, followed by box prompts increased by 10%, 20%, 30%, and 40%. Each box plot represents N=1,227 samples. Circles indicate outlier Dice scores, and the large 'X' represents the mean Dice score. No significant differences were observed between the results for different prompt sizes. The pipeline is illustrated in Figure 2.

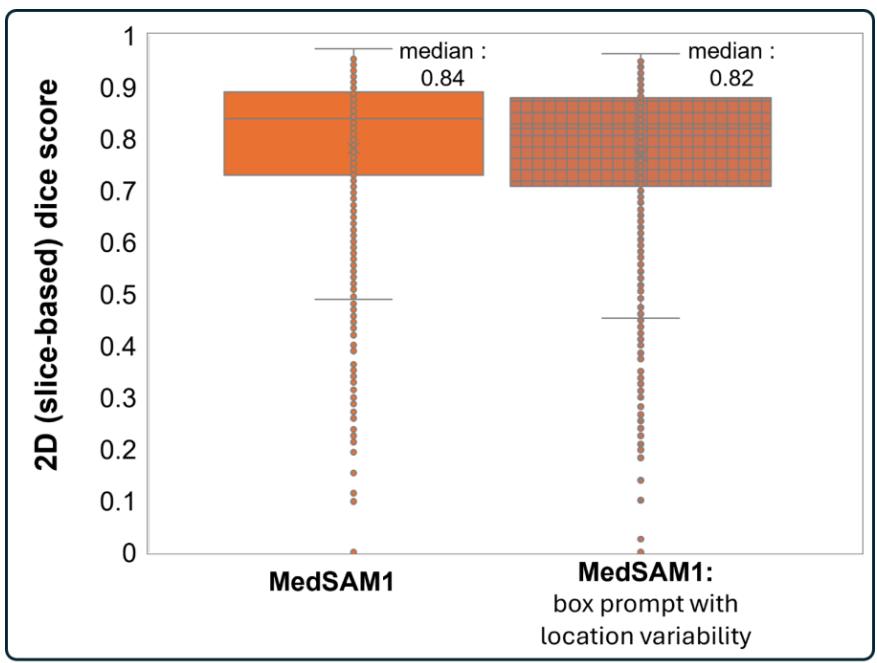

Figure 10. MedSAM1. box plot of dice scores for initial box prompt and the prompt with random movements of the box center, with a location variability of 10% of the box size, N=1,227 dice scores (each point) per box plot. Pipeline was shown in Fig 2.